\documentclass[aps,prb,twocolumn,a4paper,superscriptaddress,showpacs]{revtex4}
\usepackage{amssymb}
\usepackage{amsmath}

\usepackage[dvips]{graphicx}
\usepackage[dvips]{color}
\usepackage{subfigure}

\renewcommand{\vec}[1]{\mathbf #1}
\renewcommand{\exp}{\mathrm e}

\renewcommand{\d}{\mathrm d}

\renewcommand{\hbar}{\hslash}

\begin{document}

\title{Full counting statistics of crossed Andreev reflection}
\author{Jan Petter Morten}
\email{janpette@tf.phys.ntnu.no}
\author{Daniel Huertas-Hernando}
\affiliation{Department of Physics, Norwegian University of Science and Technology,
N-7491 Trondheim, Norway}
\author{Wolfgang Belzig}
\affiliation{University of Konstanz, Department of Physics, D-78457 Konstanz, Germany}
\author{Arne Brataas}
\affiliation{Department of Physics, Norwegian University of Science and Technology,
N-7491 Trondheim, Norway}

\date{September 17, 2008}

\begin{abstract}
We calculate the full transport counting statistics in a three-terminal
tunnel device with one superconducting source and two normal-metal or
ferromagnet drains. We obtain the transport probability distribution from
direct Andreev reflection, crossed Andreev reflection, and electron transfer
which reveals how these processes' statistics are determined by the device
conductances. The cross-correlation noise is a result of competing
contributions from crossed Andreev reflection and electron transfer, as well
as antibunching due to the Pauli exclusion principle. For spin-active tunnel
barriers that spin polarize the electron flow, crossed Andreev reflection
and electron transfer statistics exhibit different dependencies on the
magnetization configuration, and can be controlled by relative magnetization
directions and voltage bias. 
\end{abstract}

\maketitle





\date{\today}

\section{Introduction}

In crossed Andreev reflection (CA), a Cooper pair in a superconductor (S) is
converted into an electron-hole quasiparticle pair in normal-metal terminals
(N$_{n}$) or vice versa.\cite{byers:306,deutsher:apl00} This process has
potential applications in quantum information processing since it induces
spatially separated entangled electron-hole pairs. The performance of
entanglers utilizing this effect is diminished by parasitic contributions
from electron transfers between the N$_{n}$ terminals. This process will be
referred to as electron transfer (ET), but is also often denoted electron
cotunneling or elastic cotunneling.

It has been suggested that the noise properties of crossed Andreev
reflection can be used to distinguish it from electron transfer
between the normal-metal terminals.\cite{bignon:epl04} Mesoscopic
transport noise also reveals information about charge carriers which
is inaccessible through average current measurements. In a system with
several drain terminals, \emph{i.e.} current beam splitters, noise
measurements show signatures of correlations between the current flow
in separated terminals. The cross-correlation noise has attained
interest recently since it can be utilized to study
entanglement\cite{burkard:entanglementreview,martin:physletta96,anatram:prb96,torres:epjb99,gramespacher:8125,lesovik:epjb01,shechter:224513,borlin:197001,samuelsson:prl02,samuelsson:201306,taddei:134522,bayandin:prb06,lesovik:epjb01,bouchiat:77}
and correlated transport.

Crossed Andreev reflection in superconductor--normal-metal systems has been
experimentally studied in Refs. %
\onlinecite{russo:prl05,cadden:237003,beckmann:apa2007}. In Ref. %
\onlinecite{russo:prl05} the nonlocal voltage was measured in a multilayer
Al/Nb structure with tunnel contacts between the superconducting Nb and the
normal-metal Al layers. Current was injected through one of the
normal-metal--superconductor contacts, and a nonlocal voltage measured
between the superconductor and the other normal-metal. At injection bias
voltage below the Thouless energy $E_{\text{Th}}=\hslash D/d^{2}$ associated
with the separation $d$ between the normal-metals, positive nonlocal voltage
was measured and this was interpreted as the result of dominating ET. For
voltages $eV$ above $E_{\text{Th}}$ the nonlocal voltage changed sign, which
was interpreted as a consequence of dominating CA. Subsequently,
measurements reported in Ref. \onlinecite{cadden:237003} on nonlocal
voltages in Au probes connected to a wire of superconducting Al by
transparent interfaces indicated that the ET contribution is larger than the
contribution from CA.

The competition between CA and ET determines the sign of the nonlocal
voltage and has been studied theoretically using various approaches \cite%
{Falci:epl01,yamashita:prb03-174504,chtchelkatchev:jetp03,sanchez:214501,brinkman:214512,kalenkov:172503,golubev:184510,kalenkov:224506,levyyeyati:nature2007}
including the circuit theory of mesoscopic superconductivity utilized in
this paper.\cite{morten:prb06,morten:apa2007} We will consider the linear
response nonlocal conductance $G_{\text{nl}}$. In
superconductor--normal-metal hybrid devices where transport in one
normal-metal terminal N$_{1}$ is measured in response to an applied voltage
in another normal metal-terminal N$_{2}$, this quantity is defined by 
\begin{equation}
  \partial _{V_{2}}I_{1}=-G_{\text{nl}}=-(G_{\text{ET}}-G_{\text{CA}}),
  \label{eq:Gnldef}
\end{equation}
where we have introduced conductances associated with the charge
transfer processes introduced above, $G_{\text{CA}}$ for crossed
Andreev reflection, and $G_{\text{ET}}$ for electron
transfer.\cite{Falci:epl01} The sign of the nonlocal conductance is
determined by the competition between ET and CA.  Theoretical
calculations based on second order perturbation theory in the
tunneling Hamiltonian formalism predicted that the nonlocal
conductance resulting from CA reflection is equivalent in magnitude to
the contribution from ET.\cite{Falci:epl01} Thus the induced voltage
in N$_{1}$ in response to the bias on N$_{2}$ should vanish since CA
and ET give currents with opposite sign, in contrast to the
measurements reported in Refs.
\onlinecite{russo:prl05,cadden:237003,beckmann:apa2007}. The tunneling
limit was also considered in Refs.
\onlinecite{melin:174509,morten:prb06,golubev:184510}, and it has been
found that the nonlocal conductance is in fact of fourth order in the
tunneling and favours ET.

Experimental investigations of crossed Andreev reflection in
superconductor-ferromagnet (S-F) structures have been reported in Refs. %
\onlinecite{beckmann:prl04,beckmann:875}. The measurements in Ref. %
\onlinecite{beckmann:prl04} were modeled using the theory of Ref. %
\onlinecite{Falci:epl01}.

Experimental studies of the CA and ET noise properties can be used to
determine the relative contributions of these processes to the nonlocal
conductance. It was shown theoretically in  that CA contributes positively
to the noise cross correlations, whereas ET gives a negative contribution.%
\cite{bignon:epl04} Calculations of higher order noise correlators or the
noise dependence on spin-polarizing interfaces can reveal further
information about the CA and ET processes.

We will consider the full counting statistics (FCS) which encompasses all
statistical moments of the current flow.\cite%
{levitov:jetp58,belzig:NS-FCS,belzig:fcsreview} The noise properties of ET
and CA reflection thus obtained can be used to study the competition between
these processes and reveal information that is not accessible in the mean
currents. Our calculation also determines the contribution to the noise
coming from the fermion statistics (Pauli exclusion principle). Moreover,
the charge transfer probability distribution provided by FCS reveals
information about the probability of elementary processes in the circuit.%
\cite{morten-2006}

In this paper we calculate the FCS of multiterminal
superconductor-normal metal and superconductor-ferromagnet proximity
structures, and study the currents, noise and cross correlations
associated with the various transport processes. We obtain the
probability distribution for transport at one normal-metal drain, and
show that the probability associated with ET is larger than the
probability associated with CA. For spin-active interfaces we show how
spin filtering can be utilized to control the relative magnitude of
the CA and ET contributions to the transport. Finally, we consider
crossed Andreev reflection for spin triplet superconductors. This
paper goes beyond our previous publications
Refs. \onlinecite{morten:prb06,morten:apa2007,morten-2006} in that we
consider different bias voltages on the normal-metal/ferromagnetic
drains and discuss the effect of triplet superconductivity.

The paper is organized in the following way: In Sec. \ref{sec:theory} we
describe the electronic circuit and outline the formalism utilized to
calculate the cumulant generating function of the probability distributions.
In Sec. \ref{sec:normal} we discuss the results in the case of normal
metals, and in Sec. \ref{sec:ferro} we consider the spin-active connectors.
Finally, our conclusions are given in Sec. \ref{sec:conclusion}.

\section{Model\label{sec:theory}}

\begin{figure}[tbp]
\centering
\includegraphics{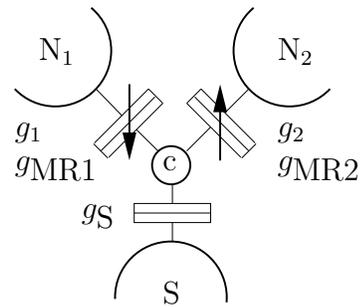}  
\caption{Circuit theory representation of the considered beamsplitter where
supercurrent flows from a superconducting reservoir (S) into normal metal
drains (N$_{n}$). Tunnel barriers between cavity (c) and the drains can be
spin-active, and are characterized by the conductances $g_{n}$ and spin
polarizations $g_{\text{MR}n}/g_{n}$.}
\label{fig:beamsplitter}
\end{figure}
The systems we have in mind can be represented by the circuit theory diagram
(see Sec. \ref{sec:ct}) shown in Fig. \ref{fig:beamsplitter}. A
superconducting source terminal (S) and normal metal drain terminals (N$_{n}$%
) are connected by tunnel barriers with conductance $g_{n}$ to a common
scattering region which is modeled as a chaotic cavity (c). The assumptions
on the cavity is that the Green function is isotropic due to diffusion or
chaotic scattering at the interfaces, and that charging effects and
dephasing can be disregarded. The tunnel barriers can be spin-active with
spin polarization $g_{\text{MR}n}/g_{n}$. We consider elastic transport at
zero temperature. The superconducting terminal is grounded, and biases $V_{n}
$ are applied to the normal terminals. We assume that $V_{n}\ll \Delta _{0}$%
, where $\Delta _{0}$ is the gap of the superconducting terminal. In
addition to the ET and CA processes described above, there can also be
direct Andreev (DA) reflection between the superconductor and one
normal-metal terminal, where both particles of the Andreev reflected pair
are transferred into N$_{n}$. Semiclassical probability arguments show that
the subgap charge current in the connector between N$_{1}$ and c in the
three terminal network in Fig. \ref{fig:beamsplitter} has the following
structure,\cite{Falci:epl01,morten:prb06} 
\begin{equation}
    \label{eq:I1}
    I_{1}(E)=\;G_{\text{CA}}(V_{1}+V_{2})-G_{\text{ET}}(V_{2}-V_{1})+2G_{\text{DA1}}V_{1},
\end{equation}
where we have introduced the conductance $G_{\text{DA1}}$ associated with
direct Andreev reflection between terminal N$_{1}$ and S. Eq. \eqref{eq:I1}
leads to the definition of the nonlocal conductance in \eqref{eq:Gnldef}
which shows that when $V_{2}>V_{1}$, ET and CA give competing negative and
positive contributions respectively to the current. The conductances in %
\eqref{eq:I1} will be determined in the calculation below.

\subsection{Circuit theory\label{sec:ct}}

The circuit theory of mesoscopic transport is reviewed in Ref. %
\onlinecite{nazarov:handbook} and is a suitable formalism to study proximity
effects in superconducting nanostructures. The theory is developed from a
discretization of the quasiclassical theory of superconductivity,\cite%
{nazarov:sm99} in combination with a theory of boundary conditions based on
scattering theory.

The circuit theory is formulated in terms of the quasiclassical Green
functions of the terminals and nodes in the system. Nodes can represent
small islands or lattice points of diffusive parts of the system. Under the
assumptions described above, the Green function of the spin-singlet S
terminal in Fig. \ref{fig:beamsplitter} is $\check{G}_\text{S}=\hat{\tau}_1$
where $\hat{\tau}_k$ is a Pauli matrix in Nambu space. The Green functions $%
\check{G}_n$ of normal-metal terminals N$_n$ are given by 
\begin{align}
\check{G}_\text{c}(E)=\begin{cases}\hat{\tau}_3\check{\tau}_3+(\check{%
\tau}_1+i\check{\tau}_2)&\vert
E\vert<eV_n,\\\hat{\tau}_3\check{\tau}_3+\text{sgn}(E)\hat{\tau}_3(\check{%
\tau}_1+i\check{\tau}_2)&\vert E\vert\geq eV_n,\end{cases}
\end{align}
where $\check{\tau}_k$ are Pauli matrices in Keldysh space.

Matrix currents $\check{I}$ describe the flow of charge, energy and
coherence between terminals and nodes through connectors, and conservation
of these currents are imposed at each node. This generalized Kirchhoff law
determines the Green function of the cavity $\check{G}_{\text{c}}$. The
balance of matrix currents $\check{I}_{n}$ flowing between each terminal $%
n=1,2,\text{S}$ and the cavity, including the effect of superconducting
pairing in c, can be written 
\begin{equation}
  \sum_{n}\check{I}_{n}-\left[ \frac{e^{2}\nu _{0}\mathcal{V}_{\text{c}}}{\hslash }\Delta _{\text{c}}\hat{\tau}_{1},\,\check{G}_{\text{c}}\right] =0.
  \label{eq:currcons}
\end{equation}
Here, $\Delta _{\text{c}}$ is the superconducting order parameter on the
cavity, $\nu _{0}$ is the density of states and $\mathcal{V}_{\text{c}}$ the
volume of the cavity.\cite{nazarov:sm99,morten:apa2007} The second term on
the left hand side of the equation above induces electron-hole pairing.
Since the pairing term has the same structure as the coupling to the
superconducting terminal (see \eqref{eq:bc}), it gives quantitative effects
that are captured by renormalizing $e^{2}\nu _{0}\mathcal{V}_{\text{c}%
}\Delta _{\text{c}}/\hslash +g_{\text{S}}\rightarrow g_{\text{S}}$ and will
be omitted in the following. Here, $g_{\text{S}}$ is the superconductor
tunnel conductance.

Spin dependent transmission and reflection is described by tunneling
amplitudes $t_{k,\sigma }^{n}$ and $r_{k,\sigma }^{n}$ for electrons with
spin $\sigma $ incident from the cavity side on the interface between the
cavity and terminal $n$ in channel $k$. The matrix current $\check{I}_{n}$
through such spin-active interfaces is \cite{Huertas-Hernando:prl02} 
\begin{align}
   \begin{split}
    \check{I}_n=&\frac{g_n}{2}\left[\check{G}_n,\check{G}_\text{c}\right]+\frac{g_{\text{MR}n}}{4}\left[\left\{\vec{m}_n\cdot\bar{\boldsymbol{\sigma}}~\bar{\tau}_3,\check{G}_n\right\},\check{G}_\text{c}\right].
  \end{split}
   \label{eq:bc}
\end{align}
Here, $g_{n}=g_{\text{Q}}\sum_{k,\sigma }|t_{k,\sigma }^{n}|^{2}$ is the
tunnel conductance where $g_{\text{Q}}=e^{2}/h$ is the conductance quantum, $%
g_{\text{MR}n}=g_{\text{Q}}\sum_{k}(|t_{k,\uparrow
}^{n}|^{2}-|t_{k,\downarrow }^{n}|^{2})$ is the conductance polarization, $%
\bar{\boldsymbol{\sigma }}$ is the vector of Pauli matrices in spin space,
and the unit vector $\vec{m}_{n}$ points in the direction of the
magnetization of the spin polarizing contact. In \eqref{eq:bc} we have
neglected an additional term related to spin dependent phase shifts from
reflection at the interface which can be suppressed by a thin, non-magnetic
oxide layer. \cite{cottet:180503,Tedrow:prl86} The effects of spin filtering
contained in the polarization $g_{\text{MR}n}$, which can be obtained
experimentally using ferromagnetic terminals, will be studied in Sec. \ref%
{sec:ferro}.

In systems where all the connectors are tunnel barriers described by the
matrix current \eqref{eq:bc}, it is possible to solve \eqref{eq:currcons}
analytically and obtain the cavity Green function in terms of the terminal
Green functions and the tunneling parameters. To this end, we note that it
is possible to write \eqref{eq:currcons} as $[\check{M},\check{G}_\text{c}]=0
$. Employing the normalization condition $\check{G}_\text{c}^2=1$, the
solution can be expressed in terms of the matrix $\check{M}$ as\cite%
{borlin:197001} 
\begin{align}
\check{G}_\text{c}=\check{M}/\sqrt{\check{M}^2}.  \label{eq:Gc}
\end{align}
This result facilitates calculation of the cumulant generating function of
the charge transfer probability distribution in tunnel barrier multiterminal
circuits.

\subsection{Full counting statistics}

Full counting statistics is a useful tool to compute currents and noise in a
multiterminal structure,\cite{nazarov:196801} and also provides the higher
statistical moments that may become experimentally accessible in these
systems. Additionally, one can obtain information about the elementary
charge transport processes by studying the probability distributions.\cite%
{morten-2006} The cumulant generating function (CGF) $\mathcal{S}(\{\chi_n\})
$ of the probability distribution is directly accessible by the Green
function method, and is defined by 
\begin{align}
  \begin{split}
    \exp^{-\mathcal{S}(\{\chi_n\})}=&\sum_{N_n}P(\{N_n\};t_0)\exp^{-i\sum_n\chi_n N_n} \\
    P(\{N_n\};t_0)=&\frac{1}{(2\pi)^M}\int_{-\pi}^{\pi}\d^M\chi\,\exp^{-\mathcal{S}(\{\chi_n\})+i\sum_n N_n \chi_n}.
    \label{eq:CGFdef}
  \end{split}%
\end{align}
Here, $P(\{N_n\};t_0)$ is the probability to transfer $N_1,\,N_2,\,...,N_n$
electrons into terminal N$_1,\,$N$_2,\,...,$N$_n$ in time $t_0$, and $M$ is
the total number of terminals in the circuit. The CGF is a function of the
set of counting fields $\{\chi_n\}$ which are embedded in the Green function
at each terminal by the transformation $\check{G}_n\to\exp^{i\chi_n\check{%
\tau}_\text{K}/2}\check{G}_n\exp^{-i\chi_n\check{\tau}_\text{K}/2}$ where $%
\check{\tau}_\text{K}=\hat{\tau}_3\check{\tau}_1$. The CGF will be
determined by the following relation,\cite{nazarov:196801} 
\begin{align}
\frac{ie}{t_0}\frac{\partial \mathcal{S}(\{\chi_n\})}{\partial \chi_n}=\int
dE~I_n(\{\chi_n\}),  \label{eq:Seq}
\end{align}
where $I_n(\{\chi_n\})$ is the particle (counting) current through connector 
$n$ in presence of the counting fields. Our task is now to integrate this
equation and obtain the CGF $\mathcal{\ S}(\{\chi_n\})$. Using the general
solution to the matrix current conservation \eqref{eq:Gc}, it was found in
Ref. \onlinecite{borlin:197001}, that this is possible by rewriting the
counting current in terms of a derivative of $\check{M}$ with respect to the
counting fields. Explicit derivation shows that 
\begin{align}
I_n(\{\chi_n\})=\frac{1}{8e}\text{Tr}\left\{ \check{\tau}_\text{K}\check{I}%
_n(\{\chi_n\})\right\}=\frac{1}{4ei}\partial_{\chi_n}\text{Tr}\left\{ \sqrt{%
\check{M}^2}\right\}.  \label{eq:IM}
\end{align}
This result is valid also in the presence of spin-active contacts %
\eqref{eq:bc}. Combining \eqref{eq:IM} with \eqref{eq:Seq} yields $\mathcal{S%
}(\{\chi_n\})$ straightforwardly.

Practical calculations of CGFs are performed by diagonalizing the matrix $%
\check{M}$, which allows us to express the CGF in terms of the eigenvalues
of $\check{M}$, 
\begin{align}
\mathcal{S}=-\frac{t_0}{4e^2}\int dE~\sum_k\sqrt{\lambda_k^2}.
\label{eq:Ssolution}
\end{align}
In this equation, $\{\lambda_k\}$ is the set of eigenvalues of $\check{M}$.

We can obtain the cumulants of the transport probability distribution by
successive derivatives of the CGF.\cite{belzig:fcsreview} Specifically, we
obtain the mean current from 
\begin{align}
I_n=-\frac{ie}{t_0}\left.\frac{\partial \mathcal{S}(\{\chi\})}{\partial
\chi_n}\right|_{\{\chi=0\}}.
\end{align}
The current noise power is given by 
\begin{align}
C_{m,n}=2\frac{e^2}{t_0}\left.\frac{\partial^2 \mathcal{S}(\{\chi\})}{%
\partial\chi_m\partial\chi_n}\right|_{\{\chi=0\}},  \label{eq:secondcumulant}
\end{align}
where in the multiterminal structure, the autocorrelation noise at terminal $%
n$ is given by $C_{n,n}$. When $m\neq n$, \eqref{eq:secondcumulant} gives
the noise cross-correlations.

\section{Normal metal drains\label{sec:normal}}

In this section, we will consider the FCS of the superconducting
beamsplitter in Fig. \ref{fig:beamsplitter} when the connectors are not spin
polarizing, and generalize previous works by taking into account a
difference in drain terminal voltages $V_1\neq V_2$.

In the regime $E<eV_{1},eV_{2}$ the only contribution to the nonlocal
conductance comes from CA since we consider zero temperature. The resulting CGF was studied in Ref. %
\onlinecite{borlin:197001} where it was assumed that $V_{1}=V_{2}$. In the
general case $V_{1}\neq V_{2}$, the total CGF $\mathcal{S}$ following from %
\eqref{eq:Ssolution} has one contribution from the energy range $%
E<eV_{1},eV_{2}$, and if the voltages are different, another contribution in
the energy range $eV_{1}\leq E<eV_{2}$ (we assume $V_{2}>V_{1}$), 
\begin{align}
\mathcal{S}& =-\frac{t_{0}}{4e^{2}}\sum_{k}\left(
\int_{-eV_{2}}^{-eV_{1}}+\int_{-eV_{1}}^{eV_{1}}+\int_{eV_{1}}^{eV_{2}}%
\right) E\sqrt{\lambda _{k}^{2}(E)}  \nonumber \\
& =\,\mathcal{S}_{\text{a}}(V_{1})+\mathcal{S}_{\text{b}}(V_{2}-V_{1}).
\label{eq:Setca}
\end{align}%
There is no contribution to transport at $|E|>eV_{2}$. Here, we have defined
two separate contributions to the CGF that govern transport in the regime $%
E<eV_{1},eV_{2}$ ($\mathcal{S}_{\text{a}}$) where only Andreev reflections
(CA and DA) can occur, and the regime $eV_{1}\leq E<eV_{2}$ ($\mathcal{S}_{%
\text{b}}$) where in addition to Andreev reflections, ET can take place. The
contribution $\mathcal{S}_{\text{a}}$ was calculated in Ref. %
\onlinecite{borlin:197001}, see \eqref{eq:S1}, where we have defined $%
g_{\Sigma }=[g_{\text{S}}^{2}+g^{2}]^{1/2}$ and $g=g_{1}+g_{2}$. The
counting factors $\exp ^{2i\chi _{\text{S}}-i\chi _{m}-i\chi _{n}}$ describe
processes where two particles are transferred from S, and one particle is
counted at terminal N$_{m}$ and at terminal N$_{n}$ ($m,n=1,2$).

\begin{widetext}
\begin{subequations}
\begin{align}
  &{\cal S}_\text{a}=-\frac{t_0 V_1}{\sqrt{2}e}\sqrt{g_\Sigma^2+\sqrt{g_\Sigma^4+4g_\text{S}^2\sum_{m,n}g_mg_n(\exp^{2i\chi_\text{S}-i\chi_m-i\chi_n}-1)}},\label{eq:S1}\\
  &S_\text{b}=-\frac{t_0(V_2-V_1)}{\sqrt{2}e}\nonumber\\
  &\times\sqrt{g_\Sigma^2+2g_1g_2(\exp^{i\chi_1-i\chi_2}-1)+\sqrt{g_\Sigma^4+4g_\text{S}^2\sum_{n}g_ng_2(\exp^{2i\chi_\text{S}-i\chi_n-i\chi_2}-1)+4g_1g_2g^2(\exp^{i\chi_1-i\chi_2}-1)}}\label{eq:S2}
\end{align}
\label{eq:Secca}
\end{subequations}
  \begin{center}
    \begin{figure}[htbp]
      \mbox{ 
        \subfigure[]{\includegraphics{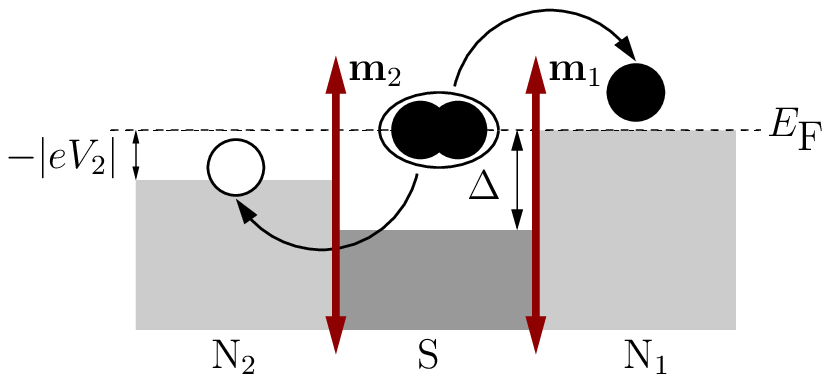}}
        \qquad
        \subfigure[]{\includegraphics{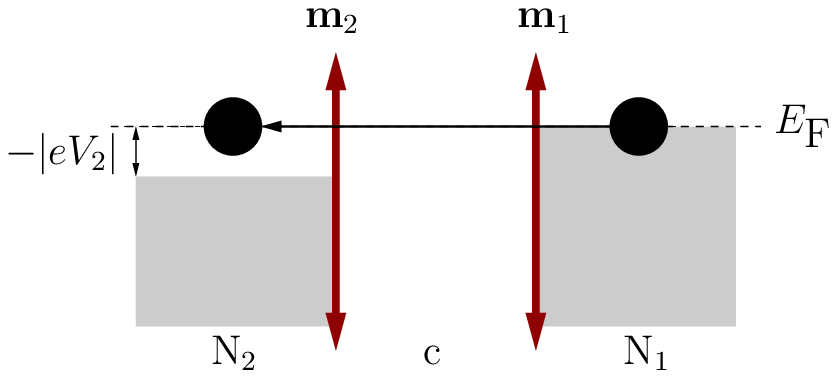}}}
      \caption{Transport processes in the three terminal device when
        $eV_1=0$: (a) Crossed Andreev reflection: A Cooper pair from S
        is converted into an electron-hole pair in c by Andreev
        reflection, and the electron with energy $+E$ is transferred
        into N$_1$, and the hole with energy $-E$ is transferred into
        N$_2$. Tunnel barriers between the reservoirs may be
        spin-active and are described by magnetization vectors
        $\vec{m}$ that in this paper are considered collinear. (b)
        Electron transfer: A particle from N$_1$ tunnels through the
        cavity c into N$_2$. The density of states in the cavity c is
        suppressed due to the proximity effect from the
        superconducting terminal.}
      \label{fig:etca}
    \end{figure}
  \end{center}
\end{widetext}


In \eqref{eq:S2} we show the calculated $\mathcal{S}_\text{b}$ which has
contributions from electron transfer. The counting factor $%
\exp^{i\chi_1-i\chi_2}$ describes events where an electron is transferred
from N$_1$ to N$_2$. Compared to $\mathcal{S}_\text{a}$, we see that DA
events between S and N$_1$ that would be described by counting factors $%
\exp^{2i\chi_\text{S}-2i\chi_1}$ no longer occur. This can be understood
from the electron-hole--nature of Cooper pairs, see Fig. \ref{fig:etca}. Two
quasiparticles, with energy $+E$ for the electron and $-E$ for the hole
constitute the Andreev reflected quasiparticle pairs in c. In the energy
range considered here, $eV_1\leq E < eV_2$, the states in N$_1$ are
occupied, precluding DA reflection into N$_1$. A similar argument shows that
in CA processes, the electron must be transferred into N$_1$ and the hole
into N$_2$.

The nonlocal conductance $G_\text{nl}=G_\text{ET}-G_\text{CA}$ following
from \eqref{eq:S2} is in agreement with Ref. \onlinecite{morten:prb06} where 
\begin{align}
G_\text{ET}=\frac{g_1g_2}{2}\frac{2g^2+g_\text{S}^2}{[g^2+g_\text{S}^2]^{3/2}%
},~G_\text{CA}=\frac{g_1g_2}{2}\frac{g_\text{S}^2}{[g^2+g_\text{S}^2]^{3/2}}.
\end{align}
The nonlocal conductance is dominated by ET and is of order $\mathcal{\ O}%
(g_n^4)$.\cite{morten:prb06,golubev:184510} When $g/g_\text{S}\ll 1$, the
nonlocal conductance vanishes due to equal probability for ET and CA as we
will explicitly show by inspection of the probability distribution below. In
the opposite limit that the coupling to normal terminals dominates, $g_\text{%
S}/g\ll 1$, the conductance for ET is to lowest order given by $%
g_1g_2(1-\delta N)/g$. Here $\delta N=(g_\text{S}/g)^2/2$ is the lowest
order correction to the low-energy density of states due to superconducting
correlations. The CA conductance becomes $g_1g_2\delta N/g$ in this limit.

Let us now consider cross-correlations. In general, CA leads to a positive
contribution and ET leads to a negative contribution to the
cross-correlation. An additional, negative contribution is induced by the
Pauli exclusion principle. The cross-correlation between N$_{1}$ and N$_{2}$
following from \eqref{eq:Setca} is 
\begin{subequations}
\begin{align}
& C_{1,2}=2e(V_{1}+V_{2})G_{\text{CA}}-2e(V_{2}-V_{1})G_{\text{ET}}
\label{eq:Cmnetca} \\
& ~-\frac{10eV_{1}}{g_{\Sigma }}(G_{1}-G_{\text{nl}})(G_{2}-G_{\text{nl}})
\label{eq:CmnPauli} \\
& ~+\frac{4e(V_{2}-V_{1})}{g_{\Sigma }}\left[ G_{\text{CA}}(G_{2}+2G_{\text{%
DA2}})-G_{\text{DA2}}G_{\text{nl}}\right] ,  \label{eq:Cmninter}
\end{align}%
where we have defined the local differential conductances $G_{n}=\partial
_{n}I_{n}$ and the conductance for direct Andreev reflection into terminal $n
$ is $G_{\text{DA}n}=g_{\text{S}}^{2}g_{n}^{2}/2g_{\Sigma }$.\cite%
{morten:prb06} We now focus on the competition between CA and ET. When the
two normal-metal terminals are at equal voltage $V_{1}=V_{2}$ the
contribution from ET and also the term in \eqref{eq:Cmninter} vanishes and
we are left with a positive contribution to the cross-correlations from CA
due to the correlated particle transfer into N$_{1}$ and N$_{2}$. An
additional, negative contribution due to the Pauli principle in %
\eqref{eq:CmnPauli} vanishes in the limit of asymmetric conductances $g_{%
\text{S}}\gg g$ or $g_{1(2)}\gg g_{2(1)},g_{\text{S}}$ due to the noisy
(Poissonian) statistics of the incoming supercurrent. A negative
contribution from ET in \eqref{eq:Cmnetca} is induced when there is a
voltage difference between the normal-metal terminals due to the currents
with opposite signs in N$_{1}$ and N$_{2}$ resulting from this process. This
demonstrates that it is possible to tune the sign of cross-correlations by
the voltages $V_{1}$ and $V_{2}$.\cite{bignon:epl04} The contribution to $C_{1,2}$ in \eqref{eq:Cmninter} is proportional to the voltage difference $V_1-V_2$, and  vanishes in the limit of asymmetric conductances  $g_{\text{S}}\gg g$ or $g_{1(2)}\gg g_{2(1)},g_{\text{S}}$.

It is interesting to compare $\mathcal{S}_\text{b}$ with the corresponding
CGF when S is in the normal state, 
\end{subequations}
\begin{align}
\mathcal{S}_\text{b}=&-\frac{t_0(V_2-V_1)}{2e}\times \\
&\sqrt{(g_1+g_\text{S}+g_2)^2+4g_2g_1\exp^{i\chi_1-i\chi_2}+4g_2g_\text{S}%
\exp^{i\chi_\text{S}-i\chi_2}}.  \nonumber
\end{align}
Here we see a contribution due to transport between N$_1$ and N$_2$ that is
similar to the one outside the double square root in \eqref{eq:S2}.
Superconductivity leads to the double square root in \eqref{eq:S2} that
takes into account the correlation of transport through c by Andreev
reflections and ET. The complicated dependence on the counting fields in %
\eqref{eq:S2} precludes a simple interpretation of $\mathcal{S}_\text{b}$ in
terms of the probabilities of elementary charge transfer processes. However,
when $g_\text{S}\gg g$ or $g_{1(2)}\gg g_{2(1)},g_\text{S}$ we can expand
the square roots in $\mathcal{S}_\text{b}$ and obtain the CGF 
\begin{align}
\mathcal{S}_\text{b}=-\frac{t_0(V_2-V_1)}{2g_\Sigma^3e}\Big[&g_1g_2(g_\text{S%
}^2+2g^2)\exp^{i\chi_1-i\chi_2}  \nonumber \\
&+g_\text{S}^2g_1g_2\exp^{2i\chi_\text{S}-i\chi_1-i\chi_2}  \nonumber \\
&+g_\text{S}^2g_2^2\exp^{2i\chi_\text{S}-2i\chi_2}\Big].
\label{eq:ec-ca-limit}
\end{align}
In this limit the CGF describes independent CA, ET, and DA Poisson
processes. The prefactors determine the average number of charges
transferred by each process in time $t_0$.

\begin{figure}[h]
\vspace{0.5cm}
\centering
\includegraphics[width=1.0\linewidth]{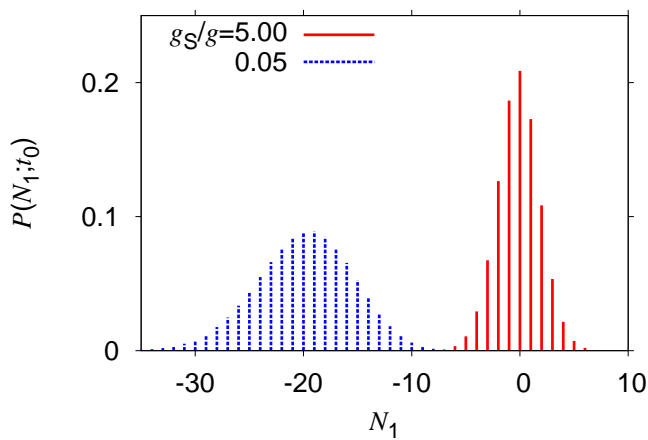}  
\caption{(Color online) Probability distribution for transport of
  $N_{1}$ electrons into terminal N$_{1}$, $P(N_{1};t_{0})$
  Eq. \eqref{eq:P-ecca}. We show distributions for two different
  values of the parameter $g_{\text{S}}/g=5.00$ (red [gray] solid
  impulses), and $0.05$ (blue [dark gray] dotted impulses). We have chosen the
  parameter $\protect\alpha =g_{1}g_{2}V_{2}t_{0}/eg=20$, which gives
  the mean value $\bar{N}_{1}$ for small $g_{\text{S}}/g$, see
  Eq. \eqref{eq:Nbar}.}
\label{fig:P-ecca}
\end{figure}

To illustrate the physics described by $\mathcal{S}_{\text{b}}$ in the limit
introduced above, let us examine the probability distribution obtained by
the definition in \eqref{eq:CGFdef}. If we consider the current response in N%
$_{1}$ to a voltage in N$_{2}$, we can consider that $V_{1}=0$ and the only
contribution to the total CGF $\mathcal{S}$ comes from $\mathcal{S}_{\text{b}%
}$. The normalized probability distribution for the transport at terminal N$%
_{1}$ following from \eqref{eq:ec-ca-limit} then becomes 
\begin{align}
P(N_{1};t_{0})=& \exp ^{-\bar{N}_{1}g_{\Sigma }^{2}/g^{2}}\sum_{\overset{%
\scriptstyle k\geq |N_{1}|}{\scriptstyle k+N_{1}\,\text{even}}}  \nonumber \\
& \times \left( \bar{N}_{1}\frac{g_{\text{S}}^{2}}{2g^{2}}\right) ^{\frac{%
k+N_{1}}{2}}\left[ \bar{N}_{1}\left( \frac{g_{\text{S}}^{2}}{2g^{2}}%
+1\right) \right] ^{\frac{k-N_{1}}{2}}  \nonumber \\
& \times \left[ \left( \frac{k+N_{1}}{2}\right) !\left( \frac{k-N_{1}}{2}%
\right) !\right] ^{-1}.  \label{eq:P-ecca}
\end{align}%
Here we have defined the mean number of particles transferred in time $t_{0}$, 
\begin{equation}
  \bar{N}_{1}=\frac{|I_{1}|t_{0}}{e}=\frac{V_{2}t_{0}}{e}\frac{g_{1}g_{2}}{g\left( 1+g_{\text{S}}^{2}/g^{2}\right) ^{3/2}}.
  \label{eq:Nbar}
\end{equation}
Eq. \eqref{eq:P-ecca} describes a joint probability distribution for CA and
ET processes with Poissonian statistics, and is constrained such that the
number of CA events described by the weight $g_{\text{S}}^{2}/2g^{2}$
subtracted by the number of ET events described by the weight $g_{\text{S}%
}^{2}/2g^{2}+1$, is $N_{1}$ as required. When $g_{\text{S}}/g\gg 1$, the
mean number of particles transferred vanishes according to \eqref{eq:Nbar}
and the probability distribution \eqref{eq:P-ecca} is symmetric around $%
N_{1}=0$. This means that the average current vanishes, since the
probabilities for ET and CA are equal. In general, the probability
distribution has its maximum for negative $N_{1}$, i.e. ET is more probable
than CA reflection. In Fig. \ref{fig:P-ecca} we have plotted the probability
distribution \eqref{eq:P-ecca} for different values of $g_{\text{S}}/g$. For
small ratios $g_{\text{S}}/g$, ET dominates and the probability distribution
is centered at a negative value for $N_{1}$. As expected, we see that the
center of the probability distribution (mean number of particles
transferred) is shifted from a negative value towards zero with increasing $%
g_{\text{S}}/g$. The width of the distribution, described by the
autocorrelation noise $C_{1,1}$ (see \eqref{eq:secondcumulant}), decreases
with increasing $g_{\text{S}}/g$. 

\section{Spin-active connectors\label{sec:ferro}}

Qualitatively, the effect of spin polarizing interfaces
on the competition between CA and ET processes in S-F systems can be
understood as follows.\cite{Falci:epl01} The ET process is favoured when magnetizations of
ferromagnetic leads are parallel since the same spin must traverse both the
interfaces between c and the ferromagnets. On the other hand, CA reflection
is favoured in an antiparallel configuration since two particles with
opposite spins must traverse the interfaces. This behaviour was
experimentally observed in Refs. \onlinecite{beckmann:prl04,beckmann:875}
where ferromagnetic Fe probes were contacted to a superconducting Al wire.

The FCS of a beam splitter with spin-active contacts and $V_1=V_2$ was
considered in our previous paper Ref. \onlinecite{morten-2006} and
constitutes $\mathcal{S}_\text{a}$, see \eqref{eq:Sagmr}. In this case the
only transport processes are DA and CA reflection, and we found that CA is
enhanced in an antiparallel alignment of the magnetizations as expected.

When $V_2 > V_1$, there can also be ET, and an additional effect is spin
accumulation at the node. With collinear magnetizations (sign of $g_{\text{MR%
}n}$ describes magnetization directions up (positive) or down (negative)
along the $z$ quantization axis), we find that in the regime $eV_1<E<eV_2$, $%
\mathcal{S}_\text{b}=\mathcal{S}_{\text{b}+}+\mathcal{\ S}_{\text{b}-}$
where $\mathcal{S}_{\text{b}\sigma}$ is given below \eqref{eq:Sbgmr}. 
\begin{widetext}
\begin{subequations}
\begin{align}
  {\cal S}_\text{a}=&-\frac{t_0 V}{\sqrt{2}e}\sqrt{1+\sqrt{1-\frac{4g_\text{MR}^2}{g_\Sigma^4}(g_\text{S}^2+g^2)+\frac{4g_\text{S}^2}{g_\Sigma^4}\sum_{m,n}\left(g_mg_n-g_{\text{MR}m}g_{\text{MR}n}\right)\left(\exp^{2i\chi-i\chi_m-i\chi_n}-1\right)}},\label{eq:Sagmr}\\
  {\cal S}_{\text{b}\sigma}=&-\frac{t_0g_\Sigma(V_2-V_1)}{2\sqrt{2}e}\left\{1+\frac{2}{g_\Sigma^2}(g_1+\sigma g_\text{MR1})(g_2+\sigma g_\text{MR2})\left(\exp^{i\chi_1-i\chi_2}-1\right)\phantom{\left [\frac{{1}^{1}}{1}\right]^{1}} \right.\nonumber\\
  &\hspace{2.8cm}+\left[1-\frac{4g_\text{MR}^2}{g_\Sigma^4}(g_\text{S}^2+g^2) + \frac{4g_\text{S}^2}{g_\Sigma^4}\sum_n(g_n-\sigma g_{\text{MR}n})(g_2+\sigma g_\text{MR2})\left(\exp^{2i\chi_\text{S}-i\chi_n-i\chi_2}-1\right)\right.\nonumber\\
  &\hspace{3.3cm}\left.\left.+\frac{4(g-g_\text{MR})^2}{g_\Sigma^4}(g_1+\sigma g_\text{MR1})(g_2+\sigma g_\text{MR2})\left(\exp^{i\chi_1-i\chi_2}-1\right)\right]^{1/2}\right\}^{1/2}.\label{eq:Sbgmr}
\end{align} 
\label{eq:S2magn}
\end{subequations}
\end{widetext}
Here we have redefined $g_\Sigma=(g_\text{S}^2+g^2+g_\text{MR}^2)^{1/2}$ and
introduced $g_\text{MR}=g_\text{MR1}+g_\text{MR2}$. The expression for $S_%
\text{b}$ reduces to the result for nonpolarizing contacts, Eq. \eqref{eq:S2}%
, in the limit that $g_{\text{MR}n}\to 0$. The two terms $\mathcal{S}_{\text{%
b}\sigma}$ correspond to the two possible directions of the spin(s) involved
in the charge transfer processes. The spin-dependent conductance for a spin
up (down) is $g_n+(-) g_{\text{MR}n}$. In ET, one spin must traverse the two
spin-active interfaces, thus the counting factor for spin $\sigma$ is
proportional to the weight $(g_1+\sigma g_\text{MR1})(g_2+\sigma g_\text{MR2}%
)$. The two spin channels are independent. The two opposite spins of an
Andreev reflected quasiparticle pair can be CA reflected into terminals with
different polarizations $g_{\text{MR}1}$ and $g_{\text{MR}2}$ according to
the prefactor $(g_1-\sigma g_\text{MR1})(g_2+\sigma g_\text{MR2})$, and each
possibility for the directions of the two spins gives an independent
contribution to the CGF $\mathcal{S}_\text{b}$.

In the limit that $g_\text{S}\gg g$ or $g_{1(2)}\gg g_{2(1)},g_\text{S}$ ($%
g_{\text{MR}n}\leq g_n$ by definition) we can expand the double square roots
and perform the summation over $\mathcal{\ S}_{\text{b}\sigma}$ which yields 
\begin{align}
\mathcal{S}_\text{b}=&-\frac{t_0(V_2-V_1)}{2eg_\Sigma^3}  \nonumber \\
&\times\left\{\left[g_\Sigma^2+(g-g_\text{MR})^2\right](g_1g_2+g_\text{MR1}g_%
\text{MR2})\exp^{i\chi_1-i\chi_2}\right.  \nonumber \\
&~~\phantom{\times}+g_\text{S}^2(g_1g_2-g_{\text{MR}1}g_\text{MR2}%
)\exp^{2i\chi_\text{S}-i\chi_1-i\chi_2}  \nonumber \\
&~~\phantom{\times}\left.+g_\text{S}^2(g_2^2-g_\text{MR2}^2)\exp^{2i\chi_%
\text{S}-2i\chi_2}\right\}.  \label{eq:SbMRexpanded}
\end{align}
The nonlocal conductance following from \eqref{eq:SbMRexpanded} is given by 
\begin{subequations}
\begin{align}
G_\text{ET}=&(g_1g_2+g_\text{MR1}g_\text{MR2})\frac{\left[g_\Sigma^2+(g-g_%
\text{MR})^2\right]}{2g_\Sigma^3}, \\
G_\text{CA}=&(g_1g_2-g_\text{MR1}g_\text{MR2})\frac{g_\text{S}^2}{2g_\Sigma^3%
}.
\end{align}
This immediately demonstrates that ET is favoured in a parallel
configuration of the magnetizations ($g_{\text{MR}1}g_{\text{MR}2}>0$) as
the same spin in this case tunnels through both interfaces. On the other
hand, CA reflection is favoured by antiparallel magnetizations ($g_{\text{MR}%
1}g_{\text{MR}2}<0$) since the opposite spins of a singlet tunnel through
different interfaces. These qualitative features are in agreement with Ref. %
\onlinecite{Falci:epl01}.

The cross-correlation following from \eqref{eq:SbMRexpanded} is 
\end{subequations}
\begin{align}
C_{1,2}=2e(V_2+V_1)G_\text{CA}-2e(V_2-V_1)G_\text{ET}.  \label{eq:C12magn}
\end{align}
The sign of $C_{1,2}$ can now be tuned by two experimental control
parameters: The bias voltages through the prefactors in \eqref{eq:C12magn},
and the relative magnetization direction that determines the magnitudes of $%
G_\text{ET}$ and $G_\text{CA}$.

In the energy range $V_1<E<V_2$ we are in this setup measuring the energy of
the quasiparticles involved in crossed Andreev reflection, see Fig. \ref%
{fig:etca}. Since the electron-like quasiparticle with energy $+E$ must flow
into N$_1$, and the $-E$ hole-like quasiparticle must flow into N$_2$, this
means that the entanglement in the energy degree of freedom of Andreev
reflected quasiparticle pairs has collapsed.

\subsection{Triplet superconductivity}

Superconducting correlations with triplet pairing symmetry in the spin
space can be induced by magnetic exchange fields in singlet
superconductor heterostructures. This effect has attained considerable
interest, and has recently been experimentally demonstrated (see
Refs. \onlinecite{buzdin:935,bergeret:1321,eschrig-review} and
references within). The different spin-space symmetry of the induced
electron-hole correlations opens interesting experimental applications
where e.g. superconducting correlations can propagate through a strong
ferromagnetic material.\cite{keizer:439,braude:077003} We have
studied the FCS when S is a source of spin triplet quasiparticle
pairs, and in this subsection we summarize our results for collinear
magnetizations when $V_{1}=V_{2}$.

The CGF for the spin triplet Cooper pairs with $S_z\left\vert
\Psi\right\rangle=0$, where $S_z$ is the spin operator along the $z$-axis
and $\left\vert \Psi\right\rangle$ is the spin part of the Cooper pair wave
function, is identical to the CGF for conventional spin singlet
superconductors. We showed in Ref. \onlinecite{morten-2006} that the CGF (%
\ref{eq:Sagmr}) reveals the entangled nature of the quasiparticle pairs. The 
$S_z\left\vert \Psi\right\rangle=0$ spin triplet states are also one of the
maximally entangled Bell states which implies that the magnetization
dependence for CA is the same for singlet and triplet in the collinear case.
This result can be shown also straightforwardly by computing the
two-electron tunneling probability $p_{1,2}$ which is proportional to $%
\left\langle \Psi\right\vert(g_1+g_\text{MR1}\bar{\sigma}_3)\otimes (g_2+g_%
\text{MR2}\bar{\sigma}_3)\left\vert \Psi\right\rangle+1\leftrightarrow 2$.

The triplet states where $S_{z}\left\vert \Psi \right\rangle =\pm \hslash $ $%
\left\vert \Psi \right\rangle $ give rise to a different dependence on the
magnetization configurations in the CGF since the quasiparticle pairs are
not in spin entangled states, but rather in product states. We obtain the CGF
\begin{widetext}
\begin{align}
  {\cal S}_{\parallel}=-\sum_\sigma\frac{t_0 V}{2\sqrt{2}e}\sqrt{1+\sqrt{1-\frac{4g_\text{MR}^2}{g_\Sigma^4}(g_\text{S}^2+g^2)+\frac{4g_\text{S}^2}{g_\Sigma^4}\left(g_{1}+\sigma g_{\text{MR1}}\right)\left(g_{2}+\sigma g_{\text{MR2}}\right)\left(\exp^{2i\chi-i\chi_m-i\chi_n}-1\right)}},\label{eq:Sagmrtrip}
\end{align}
\end{widetext}
Compared to the singlet case \eqref{eq:Sagmr}, we see that the CA
counting prefactor factorizes as a result of the non-entangled product
state of the quasiparticle pairs. This magnetization dependence can be
recovered also by calculating the two-particle tunneling probability
$p_{1,2}$ as discussed above.

\section{Conclusion\label{sec:conclusion}}

We have calculated the full counting statistics of multiterminal
tunnel-junction superconductor--normal-metal and
superconductor--ferromagnet beam splitter devices, and studied the
resulting currents and cross-correlation. The probability distribution
for transport at a normal-metal drain contact demonstrates that the
probability for electron transport between two normal-metal terminals
is larger than the probability for crossed Andreev reflection. A
finite voltage difference between the normal-metal contacts introduces
competing contributions to the cross-correlations from electron
transport between normal terminals and crossed Andreev
reflection. Finally, we have shown how spin-active contacts act as
filters for spin, and calculated the cumulant generating function. The
sign of the cross-correlation due to the competing contributions from
electron transport between drain terminals and crossed Andreev
reflection can in this case be determined by two external control
parameters, i.e. bias voltages and the relative magnetization
orientation. Finally, we make some remarks about the counting
statistics in the case of spin triplet superconductors.

\begin{acknowledgments}
This work has been supported by the the Research Council of Norway  through
Grants No. 167498/V30, 162742/V00, DFG through SFB 513 and  the
Landesstiftung Baden-W\"{u}rttemberg.
\end{acknowledgments}

\bibliography{/home/gudrun/janpette/artikkel/fs.bib}

\end{document}